\documentclass[twocolumn]{revtex4}
\usepackage{amsmath,amssymb}
\usepackage{bm}
\usepackage{epsfig}
\usepackage{color}

\def\be{\begin{equation}}
\def\ee{\end{equation}}
\def\bea{\begin{eqnarray}}
\def\eea{\end{eqnarray}}
\begin{document}
\title{Some characteristics of three exact solutions of
Einstein equations minimally coupled to a Quintessence field}

\author{Xiao-hua Zhou}
\email{xhzhou08@gmail.com}\affiliation{Department of Mathematics and
Physics, The Fourth Military Medical University, Xi'an 710032,
People's Republic of China}

\date{\today}

\begin{abstract}
We show some characteristics of three exact solutions to the
Einstein's gravity minimally coupled to a Quintessence field.
Besides eternal inflation, several other interesting inflationary
processes, such as transitory inflation, are attained in these
solutions. Singularity is avoided in some special cases.
\end{abstract}

\maketitle


\section{Introduction}
Since it has been found that the universe is undergoing an
accelerated and expanding process,$^{[1-4]}$ which is named
inflation, many methods have been put forward to explain this
phenomenon. Typical methods, the so-called dark energy models which
can be described as the Quintessence,$^{[5]}$ Phantom,$^{[6]}$
K-essence$^{[7]}$ and so on (see review in Ref.[8]), are being used
to achieve an inflationary universe. The Lagrangian of a scalar
field is minimally coupled to gravity and named Quintessence,
\bea
L=\frac{1}{2}\partial_\mu\phi\partial^\mu\phi-V(\phi),
\eea
where $V$ is a suitable potential. Under the assumption that the
universe is homogeneous and isotropic, the corresponding pressure
and density of the scalar field can be written as
\bea
p=\frac{1}{2}\dot{\phi}^2-V(\phi),\rho=\frac{1}{2}\dot{\phi}^2+V(\phi),
\eea
where an overdot denotes differential with respect to time. We
choose metric as the spatially flat
Friedman-Lemaitre-Robertson-Walker form,
$ds^2=dt^2-a^2(t)d\textbf{x}^2$, where the 3-D lien-element
$d\textbf{x}^2$ is a flat 3-D space, and take a perfect fluid
energy-momentum tensor $T^M_{\mu\nu}=(\rho_M+P_M)U_\mu U_\nu-P_M
g_{\mu\nu}$, for which the curvature scalar is
\bea
R=6\left[\frac{\ddot{a}}{a}+\left(\frac{\dot{a}}{a}\right)^2\right]=6\left(\dot{H}+2H^2\right),
\eea
where the Hubble parameter $H=\frac{\dot{a}}{a}$, and $a(t)$ is the
scale factor. The cosmological Einstein equations are
\bea
3M^2_{p}H^2-\frac{1}{2}\dot{\phi}^2-V=0,
\eea
\bea
\ddot{\phi}+3H\dot{\phi}+\frac{\partial V}{\partial\phi}=0,
\eea
where $M_{p}=(8\pi G)^{-1/2}$. The above two equations are
complicated and difficult to extract details about cosmological
evolution by solving them generally. The de Sitter solution:
$a(t)\propto \exp(Ht)~(H>0)$ and power-law solution $a(t)\propto t^A
(A>1)$ have been investigated extensively by Barrow.$^{[9-12]}$
 Besides, a complex solution with a sixth-degree potential was
obtained by Islam,$^{[13]}$ and other interesting solutions were
shown in Ref.[14-17]

Generally speaking, inflationary solution, which is important to
avoid the flatness, horizon and isotropic problems, means the
accelerated and expanding universe, in which, we need
\bea
\ddot{a}(t)>0,
\eea
where one should note that the expanding universe condition $H>0$
should be satisfied in the meantime. This condition is easily to be
satisfied with the de Sitter solution and power-law solution. We
suppose that condition (6) is satisfied in a region:
\bea
0\leq t_b\leq t\leq t_e,
\eea
In inequality (7), if $ t_b>0$, we name this process the transitory
inflation; if $t_b=0$, we name it the immediate inflation because
inflation occurs with the beginning of the universe immediately; and
if $t_b=0$ and $t_e\rightarrow\infty$, we name it the eternal
inflation. In this letter, we discuss three exact solutions which
contain several interesting inflationary processes.

\section{Three exact solutions}

\subsection{Solution one}
\bea
a(t)=a_0 \exp\left[-\alpha t\ln\left(\frac{t}{t_0}\right)+\beta
t\right],
\eea
\bea
\phi=\pm2\sqrt{2\alpha} M_{p}\sqrt{t}+\phi_0,
\eea
\bea
\nonumber V(\phi) &=&
                      M_{p}^2\alpha^2\bigg\{3\left[1-\frac{\beta}{\alpha}+2\ln\left(\frac{\phi-\phi_0}{\phi_p
                          }\right)\right]^2\\
                  & & -\frac{8M_{p}^2}{(\phi-\phi_0)^2}\bigg\},
\eea
where $a_0>0,\alpha\geq0,t_0>0$ and $\beta,\phi_0$ are constants and
$\phi_p=2M_{p}\sqrt{2\alpha t_0}$. A simple method to prove that
\emph{solution one} and the following two solutions are indeed
solutions of Eqs.(4) and (5) is shown in appendix A. If $\alpha=0$,
this solution is reduced to be de Sitter (or anti-de Sitter)
solution. Moreover, several characteristics of this solution can be
found.

\emph{(1). Expanding process}. The Hubble parameter is
\bea
H=\beta-\alpha\left[1+\ln\left(\frac{t}{t_0}\right)\right].
\eea
Let $\xi=e^{-1+\beta/\alpha}$, it is easy to find that expanding
universe condition $H>0$ induces
\bea
t<\xi t_0.
\eea
At $t=\xi t_0$, the universe reaches its biggest extension
$a_{\max}=a_0e^{\beta t_0\xi}\xi^{-\alpha t_0\xi}$.

 \emph{(2). Inflationary process}. Inflationary
condition $\ddot{a}>0$ leads to
\bea
t\ln^2\left(\frac{t}{\xi t_0}\right)-\frac{1}{\alpha}>0.
\eea
Now, we introduce two new parameters $x=t(\xi t_0)^{-1}$ and
$\eta=(\alpha\xi t_0)^{-1}$, then one can find that condition (13)
will be reduced to a simple state in which there are only two
parameters
\bea
x\ln^2(x)>\eta.
\eea
Two parameters make sure that a general picture of inequality (14)
can be shown in a planar figure. We let $f(x)=x\ln^2(x)$ and show it
in Fig. 1. It is easy to find that $f(0)=f(1)=0$ and $f(x)$ reaches
its local maximum $f_{\max}=4e^{-2}$ at $x=e^{-2}$. Noting that
expanding condition $H>0$ satisfies $t<\xi t_0 (x<1)$ and $f(x)$
reaches its minimum at $t=\xi t_0 (x=1)$, thus the existence of
inflation requires $\eta<4e^{-2}$. Let $x_1,x_2,x_3 (x_1< x_2< x_3)$
be the three solutions of $f(x)=\eta$ with $\eta<4e^{-2}$,
transitory inflation occurs only in the range $x_1<x<x_2~(x_1\xi
t_0<t<x_2\xi t_0)$. Whether $\eta<4e^{-2}$ or $\eta\geq4e^{-2}$, the
universe will change from expanding to collapsing when it crosses
$t=\xi t_0$. Therefore, this solution provides us with an
interesting evolution process which is different to de Sitter
solution and power-law solution, in which inflation occurs
immediately and forever.
 \begin{figure}
\includegraphics[scale =0.52]{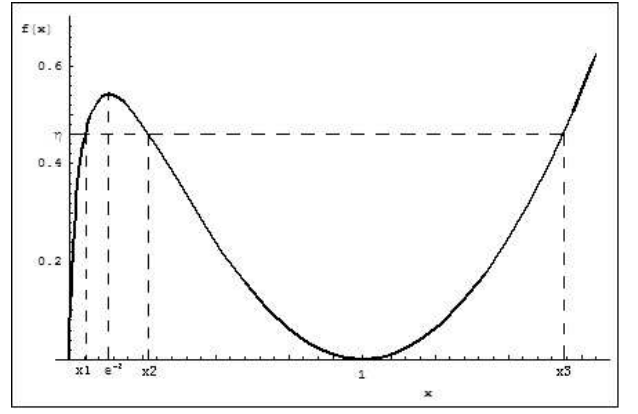}
\caption{In this chart, $x_1, x_2$ and $x_3$ are the three solutions
of $f(x)=\eta$ with $\eta<4e^{-2}$. Expanding universe satisfies
$x<1~(t<\xi t_0)$ and transitory inflation occurs in the range
$x_1<x<x_2~(x_1\xi t_0<t<x_2\xi t_0)$.}
\end{figure}

\emph{(3). Singularity}. Submitting (9) into (3), the curvature
scalar is
\bea
R=6\alpha\left[2\alpha\ln^2\left(\frac{t}{\xi
t_0}\right)-\frac{1}{t}\right].
\eea
To locate singularities in this solution, we should calculate the
Kretschmann scalar $K=R_{abcd}R^{abcd}$, which is
\bea
\nonumber
K &=& 24\bigg\{\left[\alpha-\beta+\alpha\ln\left(\frac{t}{t_0}\right)\right]^4\\
  & & -\frac{\alpha}{t}\left[\alpha-\beta+\alpha\ln\left(\frac{t}{t_0}\right)\right]^2+\frac{\alpha^2}{2t^2}\bigg\}.
\eea
It indicates that there are two singularities in this solution. One
is at the initial time $t=0$ and the other will occur in
 the infinite future.

\emph{(4). State parameter}. The state parameter
\bea
\omega=\frac{P}{\rho}=-1+\frac{2\eta}{3 x \ln^2 x},
\eea
it reaches its minimum
\bea
 \omega_{\min}=-1+\frac{e^2\eta}{6}
\eea
 at $x=e^{-2}$. Nearby $x=e^{-2}$, it is just the  inflation
 process (see Fig. 1).

\subsection{Solution two}

\bea
a(t)=a_0 \exp\left(\mu t-\lambda t^n\right),
\eea
\bea
\phi=\pm\frac{2}{n}\sqrt{2n(n-1)\lambda}M_p t^{\frac{n}{2}}+\phi_0,
\eea
\bea
\nonumber
 V(\phi) &=& 3M_{p}^2\left\{\mu-2^{\frac{3}{n}-3}n\lambda\left[\frac{n(\phi-\phi_0)}{\phi_p}\right]^{2-\frac{2}{n}}\right\}^2 \\
         & & -2^{\frac{6}{n}-3}(n-1)n\lambda M_{p}^2\left[\frac{n(\phi-\phi_0)}{\phi_p}\right]^{2-\frac{4}{n}},
\eea
where $a_0>0,\mu,\lambda,\phi_0$ are constants, and where
$\phi_p=4M_{p}\sqrt{(n-1)n\lambda}$. This solution is given by
Barrow and Liddle in Ref.[18]. In (20), we need
$(n-1)n\lambda\geq0$, which leads to the following conditions
\bea
& & \lambda>0,~n<0;\\
& &  \lambda>0,~n>1;\\
& &  \lambda<0,~0< n<1;\\
& &  (n-1)n\lambda=0.
\eea
 Several characteristics of this solution are shown as below.

\emph{(1). De Sitter solution and anti-de Sitter solution.} This
solution can also be reduced to de Sitter solution in the following
two cases
\bea
& &  \mu>0,~n\lambda=0;\\
& &  \mu>\lambda,~n=1.
\eea
Anti-de Sitter solution can be attained in two state
\bea
 & & \mu<0,~n\lambda=0;\\
 & & \mu<\lambda,~n=1.
\eea

 \emph{(2). Expanding process}. The Hubble
parameter is
\bea
H=-n \lambda t^{n-1}+\mu.
\eea
Considering conditions in (22)-(25), expanding universe condition
$H>0$ is valid in the following cases
\bea
& & t\geq0~for~\lambda>0,n<0,\mu\geq0;\\
& & t<\left(\frac{\mu}{n\lambda}\right)^{\frac{1}{n-1}}~for~\lambda>0,n<0,\mu<0;\\
& & t<\left(\frac{\mu}{n\lambda}\right)^{\frac{1}{n-1}}~for~\lambda>0,n>1,\mu>0;\\
& & t\geq0~for~\lambda<0,0<n< 1,\mu\geq0;\\
& & t<\left(\frac{\mu}{n\lambda}\right)^{\frac{1}{n-1}}~for~\lambda<0,0<n<1,\mu<0;\\
& & t\geq0~for~\mu>0,~n\lambda=0;\\
& & t\geq0~for~\mu>\lambda,~n=1.
\eea
In the above cases, (31), (34), (36) and (37) are the eternal
expanding states, and the universe will turn from expanding to
collapsing at $t=\left(\frac{\mu}{n\lambda}\right)^{\frac{1}{n-1}}$
in the other states.

\emph{(3). Inflationary process}. Inflationary condition
$\ddot{a}>0$ is reduced to
\bea
n^2\lambda^2t^{2n}+\mu^2t^2-n\lambda t^n (2\mu t+n-1)>0.
\eea
Unlike \emph{solution one}, many constants make it difficult to give
a general picture of the valid region of condition (38). We will
give an example in the later part of this section, in which
inflation condition is expressed in apparent state.

\emph{(4). Singularity}. The curvature scalar is
\bea
R=6\left[(1-n)n\lambda t^{n-2}+2\left(\mu-n\lambda
t^{n-1}\right)\right],
\eea
and the Kretschmann scalar is
\bea
\nonumber K &=& 12\left\{\left[\left(n\lambda t^{n-1}\right)^2+\mu^2-n\lambda t^{n-2}(2\mu t+n-1)\right]^2\right.\\
            & & +\left.\left(\mu-n\lambda t^{n-1}\right)^4\right\}.
\eea
If $n<1$, the universe begins with an initial singularity and it
will evolve to de Sitter (or anti-de Sitter) phase as
$H\simeq\mu(R\simeq12\mu)$ when $t\rightarrow\infty$. Specially,
$n=1$ is the de Sitter (or anti-de Sitter) spacetime. Else if
$1<n<2$ the universe begins with an initial singularity and it will
evolve to another end-singularity. If $n\geq2$, the universe begins
with de Sitter spacetime and it will evolve to a singularity in the
infinite future.

Now, let's see an example for $n=-1$. Then this solution can be
written as
\bea
a(t)=a_0 \exp\left(\mu t-\frac{\lambda}{t}\right),
\eea
\bea
\phi=\pm4M_p\sqrt{\lambda/t}+\phi_0,
\eea
\bea
\nonumber
  V(\phi)=M_{p}^2\Bigg\{3\left[\mu+\left(\frac{\phi-\phi_0}{\phi_p}\right)^4\right]^2-\left(\frac{\phi-\phi_0}{\phi_p}\right)^6\Bigg\},\\
\eea
 where $\lambda\geq0$. Several characteristics of this solution can be found as
follows.

\emph{(1) Expanding process}. The Hubble parameter is
\bea
H=\mu+\frac{\lambda}{t^2}.
\eea
For $n=-1$, valid expanding universe conditions (31), (32) and (36)
can be simplified as
\bea
& & t\geq0~for~\lambda>0,\mu\geq0;\\
& & t<\sqrt{-\lambda/\mu}~for~\lambda>0,\mu<0;\\
& & t\geq0~for~\lambda=0,\mu>0.
\eea

 \emph{(2) Inflationary process}. Condition $\ddot{a}>0$ in
(38) can be simplified as
\bea
(\mu t^2+\lambda)^2-2\lambda t>0.
\eea
 The left hand of inequality (48)
is a high-order function of $t$, and which is difficult to be solved
generally. However, by letting $x=\mu t^2/\lambda$ and
$\delta=\frac{4}{\mu\lambda}$ ($\mu\lambda\neq0$) , condition (48)
can be reduced to the following simple forms in which there are only
two parameters
\bea
& & (1+x)^4/x>\delta~for~\mu>0;\\
& &  (1+x)^4/x<\delta~for~\mu<0.
\eea
Specially, when $\mu\lambda=0$, we have
\bea
& & t<\lambda/2~for~\mu=0,\lambda>0;\\
& & t\geq0~for~\mu>0,\lambda=0.
\eea
Noting that expanding conditions in (45)-(47) should be satisfied in
the meantime, allying the above conditions (45)-(47) with (49)-(52),
one gets the valid inflationary regions
\bea
& & t<\lambda/2~for~\mu=0,\lambda>0;\\
& & t\geq0~for~\mu>0,\lambda=0;\\
& & (1+x)^4/x>\delta~for~\mu>0,\lambda>0;\\
& & (1+x)^4/x<\delta~and~x>-1~for~\mu<0,\lambda>0.
\eea
In the above cases, (53) is immediate inflationary state and (54) is
de Sitter inflationary process. Let $f(x)=(1+x)^4/x$, a general
picture of conditions (55) and (56) can be obtained, which is shown
in Fig. 2. One can find that $f(x)$ has a local minimum
$f_{\min}=\frac{256}{27}$ at $x=\frac{1}{3}$ and a local maximum
$f_{\max}=0$ at $x=-1$. When $\delta>0~(\mu>0)$, if
$\delta<\frac{256}{27}~(\mu\lambda>\frac{27}{64})$ condition (55) is
always satisfied and inflation will undergo forever; if
$\delta\geq\frac{256}{27}~(\mu\lambda\leq\frac{27}{64})$, there are
two inflationary processes: the immediate inflation when $x<x_1~
(t<\sqrt{\lambda x_1/\mu})$ and the late-time inflation when $x>x_2~
(t>\sqrt{\lambda x_2/\mu})$, where $x_1, x_2~(x_1\leq x_2)$ are the
two solutions for $f(x)=\delta$. When $\delta<0~(\mu<0)$, letting
$x_3,x_4 ~(x_3<x_4)$ be the two solutions of $f(x)=\delta$,
considering condition (56), one finds that there is only one
inflationary process: the immediate inflation when
$x>x_4~(t<\sqrt{\lambda x_4/\mu})$.

\begin{figure}
\includegraphics[scale =0.55]{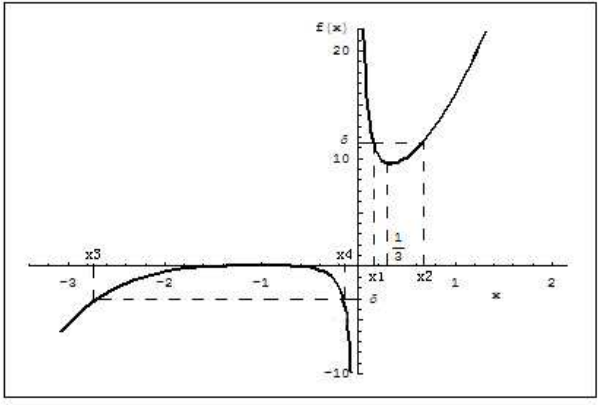}
\caption{In this chart, $x_1$ and $x_2$ are the two solutions of
$f(x)=\delta$ with $\delta\geq256/27~(\mu\lambda\leq\frac{27}{64})$,
while $x_3$ and $x_4$ are the two solutions of $f(x)=\delta$ with
$\delta<0~(\mu<0)$. Immediate inflation occurs when $x<x_1~
(t<\sqrt{\lambda x_1/\mu})$ for $\delta\geq256/27$ and when
$x>x_4~(t<\sqrt{\lambda x_4/\mu})$ for $\delta<0$; late-time
inflation occurs when $x>x_2~ (t>\sqrt{\lambda x_2/\mu})$ for
$\delta\geq256/27$.}
\end{figure}

\emph{(3). State parameter}. The corresponding state parameter is
\bea
\omega=-1+\frac{2\sqrt{\delta x}}{3(1+x)^2}.
\eea
If $\delta>0$ ($\mu>0$), one finds that $\omega$ reaches its maximum
\bea
\omega_{\max}=-1+\frac{\sqrt{3\delta}}{8}
\eea
at $x=\frac{1}{3}~(t=\sqrt{\lambda/(3\mu)})$. If $\delta<0$
($\mu<0$), one gets the state parameter $\omega\rightarrow\infty$
when $x\rightarrow -1~(t\rightarrow \sqrt{-\lambda/\mu})$. Specially
when $\mu=0$, one can easily find that $\omega$ increases linearly
with the evolvement of $t$. We now turn to investigate the
evolvement of the potential in Eq.(43). By solving
$\frac{dV}{d\phi}=0$, we obtain
\bea
\phi=\phi_0,\phi=\phi_0\pm2M_p\sqrt{1\pm\sqrt{1-16\lambda\mu}}.
\eea
Choosing suitable constants, we will get several different
evolvement of the potential. Fig. 3 shows an example of $V(\phi)$ in
Eq.(43) with $0<16\lambda\mu<1$. From it We can see that the curve
has three turning points at $\phi=\phi_0, \phi=\phi_1$ and
$\phi=\phi_2$. The universe begins with $\phi=\phi_0$ and clams up
to the local maximum of $V(\phi)$ at which $\phi$ reaches $\phi_1$.
After that, the universe will roll down from the maximum of
$V(\phi)$ to the minimum of $V(\phi)$ at $\phi=\phi_2$ or it will
roll back to $\phi_0$. This process is close to the one shown in
Ref.[13].
\begin{figure}
\includegraphics[scale =0.46]{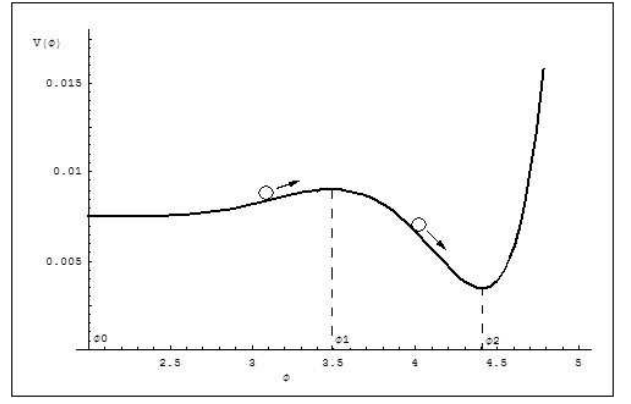}
\caption{The chart of $V(\phi)$ given by (43) with
$M_p=1,\mu=0.005,\lambda=1$ and $\phi_0=2$. The curve has three
turning points at $\phi=\phi_0, \phi=\phi_1$ and $\phi=\phi_2$. This
process is close to the one shown in Ref.[13].}
\end{figure}
\subsection{Solution three}
\bea
a(t)=\sigma\left( e^{\mu t}-\tau e^{-\mu t}\right)^n,
\eea
where $\sigma>0,\tau,\mu,n$ are suitable constants, and we need
\bea
& & \mu=any~value, \tau\leq0,n\leq0;\\
& & \mu>0,0<\tau\leq1,n>0.
\eea
Here, $(\mu=any~value, \tau\leq0)$ or $(\mu>0,0<\tau\leq1)$ will
make sure $e^{\mu t}-\tau e^{-\mu t}\geq0$ in (60) in the region
$0\leq t<\infty$. For $0<\tau\leq1$ and $n>0$, we have
\bea
\phi=\pm\sqrt{2n}M_p\log\left(\frac{e^{\mu t}-\sqrt{\tau}}{e^{\mu
t}+\sqrt{\tau}}\right)+\phi_0,
\eea
\bea
\nonumber
V(\phi)=\frac{n}{2}\mu^2M_{p}^2\Bigg[1+3n+(3n-1)\cosh\left(\frac{\phi-\phi_0}{\phi_p}\right)\Bigg].
\\
\eea
In this case, the state parameter
\bea
\omega=-1+\frac{8\tau e^{2\mu t}}{3n(\tau+e^{2\mu t})^2}.
\eea
It reaches its maximum $\omega_{\max}=-1+\frac{2}{3n}$ at
$t=\frac{\ln{\tau}}{2\mu}$. For $\tau<0$ and $n<0$, we attain
\bea
\phi=\pm2\sqrt{-2n}M_p \arctan\left(e^{\mu
t}/\sqrt{-\tau}\right)+\phi_0,
\eea
\bea
\nonumber
V(\phi)=\frac{n}{2}\mu^2M_{p}^2\Bigg[1+3n+(3n-1)\cos\left(\frac{\phi-\phi_0}{\phi_p}\right)\Bigg].\\
\eea
Then the corresponding state parameter is
\bea
\omega=-1+\frac{8\tau e^{2\mu t}}{e^{4\mu t}-3(n-1)e^{2\mu t}+3n
\tau^2}.
\eea
Here in (64) and (67), $\phi_0$ and $\phi_p$ are constants and we
need $\phi_p=M_{p}2^{\frac{-|n|}{2n}}\sqrt{|n|}$. Apparently, de
Sitter (or anti-de Sitter) solution is obtained if $\tau=0$.
Moreover, this solution reduces to the `$\sinh$' solution if
$(\tau=1,n=1)$ and to the `$\cosh$' solution if $(\tau=-1,n=1)$
$^{[14]}$ (there is a small difference, we take the flat metric, but
it is not in Ref.[14]). Actually, the potentials in (64) and (67)
have been obtained by Spalinski.$^{[16]}$ Let's see several
characteristics of this solution.

\emph{(1). Expanding process}. The Hubble parameter is
\bea
H=\frac{n\mu\left(e^{\mu t}+\tau e^{-\mu t}\right)}{e^{\mu t}-\tau
e^{-\mu t}}.
\eea
By allying (61) with (62) and noting $e^{\mu t}-\tau e^{-\mu t}>0$,
the valid regions of expanding condition $H>$ are
\bea
& & t\geq0~for~\mu>0,0\leq\tau\leq1,n>0;\\
& & t<\frac{\ln(-\tau)}{2\mu}~for~\mu>0,\tau<-1,n<0;\\
& & t<\frac{\ln(-\tau)}{2\mu}~for~\mu<0,-1<\tau<0,n<0.
\eea
In case (70), the universe will expand forever. In cases (71) and
(72), expanding process stops and the universe will turn to collapse
at $t=\frac{\ln(-\tau)}{2\mu}$.

\emph{(2). Inflationary process}. $\ddot{a}>0$ induces
\bea
\left(e^{\mu t}+\tau e^{-\mu t}\right)^2-\frac{4\tau}{n}>0.
\eea
Noting that $n\tau\geq0$ in conditions (61) and (62), the above
inequality is changed into the following two cases
\bea
& & e^{2\mu t}-2\sqrt{\tau/n}e^{\mu t}+\tau>0;\\
& & e^{2\mu t}+2\sqrt{\tau/n}e^{\mu t}+\tau<0.
\eea
Although the above two inequations are linear after the parameter
transformation: $x=e^{\mu t}$, considering expanding conditions in
(70)-(72), many constants make it difficult to find the valid
inflationary regions. We will show an example in the later part of
this section.

\emph{(3). Singularity}. The curvature scalar is
\bea
R=\frac{12n\mu^2\left[n e^{4\mu t}+2\tau(n-1)e^{2\mu
t}+n\tau^2\right]}{\left(e^{2\mu t}-\tau\right)^2},
\eea
Specially, if $n=\frac{1}{2}$, the curvature scalar is a constant
$R=3\mu^2$. The Kretschmann scalar is
\bea
\nonumber K &=& 24n^2\mu^4(e^{2\mu t}-\tau)^{-4}\Big[n^2e^{8\mu
                t}+4n\tau(n-1)e^{6\mu t}\\\nonumber
            & & +2(3n^2-4n+4)\tau^2e^{4\mu t}\\
            & & +4n(n-1)\tau^3e^{2\mu t}+n^2\tau^4\Big].
\eea
Considering the conditions in (61) and (62), if $(\mu=any~value,
\tau<0)$ or $(\mu>0,0<\tau<1)$, there is no singularity in this
solution; if ($\mu>0,\tau=1$), the universe begins with an initial
singularity and will evolve into de Sitter spacetime as $H\simeq n
\mu$ and $R\simeq 12n^2\mu^2$ when $t\rightarrow\infty$. Specially,
$\tau=0$ is de Sitter (or anti-de Sitter) spacetime.

Now we give an example for $n=1$. Then the solution is reduced to
\bea
& & a(t)=\sigma\left( e^{\mu t}-\tau e^{-\mu t}\right),\\
& & \phi=\pm\sqrt{2}M_p\log\left(\frac{e^{\mu t}-\sqrt{\tau}}{e^{\mu
     t}+\sqrt{\tau}}\right)+\phi_0,\\
& &
V(\phi)=\mu^2M_{p}^2\left[2+\cosh\left(\frac{\sqrt{2}(\phi-\phi_0)}{M_p}\right)\right].
\eea
For $n=1$, condition (62) is simplified as
\bea
\mu>0,0<\tau\leq1.
\eea
The valid expanding region is only (70)
\bea
t\geq0,0<\tau\leq1,\mu>0.
\eea
The above conditions in (81) and (82) indicate that the universe
will expand forever. Inflationary condition (74) can be written as
\bea
(e^{\mu t}-\sqrt{\tau})^2>0.
\eea
Clearly, inflation will continue forever in this case.

\section{Conclusions}

In summary, the non-linear flow equation is a widely-used approach
to explore the inflationary behavior of the universe, and our
discussion provide us with several interesting evolutionary models
of the universe. By our discussion, the fate of the universe depends
strongly on those constants, different constants are corresponding
to different cosmological evolvements. The most prominent
characteristics of those solutions are that inflation will not
undergo forever in some spacial cases, which makes those solutions
that need to be studied deeply because current investigation do not
support typical interne inflation.$^{[19]}$

In Ref.[20], Parsons and Barrow give a way to generate a family of
new exact solutions. In their method, if there is a solution
$a(t)=\exp[f(t)]$ ($f(t)$ is an arbitrary function), we can get
another exact solution with the form $a(t)=\exp[At+f(t)]$.
Specifically , by choosing $f(t)=-\alpha t\ln(t/t_0)$, (note that
$a(t)=\exp[-\alpha t\ln(t/t_0)]$ is an exact solution)\emph, we
obtain \emph{solution one}. For \emph{solution two}, there is
$f(t)=-\lambda t^n$, which is corresponding to the intermediate
inflationary: $a(t)=\exp(At^n),~0<n<1$ in Refs.[21-23]. In this
letter, we present the full valid regions of those constants and
give some useful information about this solution. However,
\emph{solution three} cannot be obtained by this way. It can be
written as $a(t)=\exp[\ln\delta+n\mu t+n \ln(1-\tau e^{-2\mu t})]$,
and it is easy to find that $a(t)=\exp[n \ln(1-\tau e^{-2\mu t})]$
is not an exact solution.

 Recently, Spalinski finds a way,$^{[16]}$ which is different
to the standard truncating method,$^{[24]}$ to attain exact
solution. In his solutions, the Hubble parameter $H$ and potential
$V$ are the function of $\phi$, but it is difficult to get the
scalar factor $a(t)$ with $t$ as a variable. In this paper,
\emph{solution three} gives us an analytical function of $a(t)$ in
(60) and the corresponding $H(t)$ in (69). Moreover, in this
solution, the periodic potential in (67) is unlike the other
potentials in (10), (21) and in (64) which will go to infinite value
when $\phi=\phi_0$ or $\phi\rightarrow\infty$, and singularity is
avoided in some special cases, all of which leads it to be a
recommendable solution.

\section{Acknowledgments}

We thank Prof. Barrow for his earnest help and useful suggestions.
We also want to express our thanks to Wang Jing for her helpful
correction of our manuscript.

\appendix{}

\section {}
  In this appendix, we give a simple way to prove that \emph{solution one} is indeed a solution of Eqs.(4) and
  (5). The other two solutions in this paper can be tested by this way.

  Taking Eq.(4) with a differential of $t$, allying Eq.(5), one
  gets
$$
 \phi = \pm M_p\int\sqrt{-2\dot{H}}dt,\eqno{(84)}
 $$
 $$
  V = M_{p}^2(3H+\dot{H}),\eqno{(85)}
$$
 where one should note $H=\frac{\dot{a}}{a}$ and $\dot{H}=\frac{\ddot{a}a-\dot{a}^2}{a^2}$.
 In general cases, we suppose that the primitive function of the right side
of Eq.(84) is existent for a given function $a(t)$, and we get an
analytical function $\phi=f(t)$ with $t$ as the variable. Then we
need to attain the inverse solution of $\phi=f(t)$:

$$
t=f^{-1}(\phi),\eqno{(86)}
$$

where the superscript $-1$ denotes to the corresponding inverse
function. Submitting (86) into (85), one will get $V$ with $\phi$ as
the variable. The above process can be used to validate the three
solutions in this paper. Let's see an example. Inserting (8) into
Eq. (84), one attains Eq.(9). Then one gets the inverse solution of
Eq.(9)
$$
t=\frac{1}{8\alpha}M_{p}^{-2}(\phi-\phi_0)^2.\eqno{(87)}
$$
Inserting (8) into Eq.(85), one has
$$
V=M_{p}^{2}\left\{3\left[\alpha-\beta+\alpha\ln\left(\frac{t}{t_0}\right)\right]^2-\frac{\alpha}{t}\right\}.\eqno{(88)}
$$
Submitting (87) into (88), one will attain $V$ with $\phi$ as the
variable in (10).

\end{document}